# Enhanced X-ray-phase-contrast-tomography brings new clarity to the 2000-year-old "voice" of Epicurean philosopher Philodemus

I. Bukreeva[1,2], A. Mittone[3], A. Bravin[3], G. Festa[4,5,6], M. Alessandrelli[7], P. Coan[3,8], V. Formoso[9,10], R. G. Agostino[9,10], M. Giocondo[9], F. Ciuchi[9], M. Fratini[1], L. Massimi[1], A. Lamarra[7], C. Andreani[4,5,11], R. Bartolino[9,10,12], G. Gigli[13], G. Ranocchia[7]*, A. Cedola[1]*

Corresponding authors to: alessia.cedola@cnr.it; graziano.ranocchia@cnr.it

[1] Consiglio Nazionale delle Ricerche, Istituto di Nanotecnologia, Rome Unit, I-00195 Rome, Italy
[2] P. N. Lebedev Physical Institute, Russian Academy of Sciences, Leninskii pr., 53 Moscow, Russia
[3] European Synchrotron Radiation Facility, F-38043 Grenoble, Cedex 9, France
[4] Università degli Studi di Roma Tor Vergata, Dipartimento di Fisica, I-00133 Rome, Italy
[5] Consiglio Nazionale delle Ricerche, Istituto di Cristallografia (Bari), I-70126 Bari, Italy
[6] Museo Storico della Fisica e Centro Studi e Ricerche Enrico Fermi, I-00184 Rome, Italy
[7] Consiglio Nazionale delle Ricerche, Istituto per il Lessico Intellettuale Europeo e Storia delle Idee, I-00161 Rome, Italy
[8] Ludwig-Maximilians-Universität, Faculty of Medicine and Department of Physics, D-80799 München, Germany
[9] Consiglio Nazionale delle Ricerche, Istituto di Nanotecnologia, Cosenza Unit, I-87036 Arcavacata Di Rende (Cosenza), Italy
[10] Università della Calabria, Dipartimento di Fisica, I-87036 Arcavacata Di Rende (Cosenza), Italy
[11] Consiglio Nazionale delle Ricerche, Istituto per i Processi Chimico Fisici (Messina), I-98158, Italy
[12] On leave to Interdisciplinary Center B. Segre Accademia Nazionale dei Lincei, I-001564 Rome, Italy
[13] Consiglio Nazionale delle Ricerche, Istituto di Nanotecnologia, I-73100 Lecce, Italy

A collection of more than 1800 carbonized Greek and Latin papyri, discovered in the Roman 'Villa dei Papiri' at Herculaneum in the middle of 18th century, is the unique classical library survived from antiquity. These ancient-Herculaneum-papyri were charred during 79 A.D. Vesuvius eruption, a circumstance which providentially preserved them until now. This magnificent collection contains valuable work by Greek philosophers, such as Epicurus, Chrysippus and Philodemus, in particular an impressive amount of extensive treatises by Philodemus of Gadara, an Epicurean philosopher of the 1st century BC[1,2].

The aim of the present study is to read extended and hitherto unknown portions of text hidden inside carbonized-Herculaneum-papyri using enhanced X-ray-phase-contrast-tomography (XPCT) non-destructive technique[3,4] and a new set of numerical algorithms for 'virtual-unrolling'.

This paper documents our success in revealing the largest portion of Greek text ever detected so far inside unopened scrolls, with unprecedented spatial resolution and contrast, all without damaging these precious historical manuscripts. Parts of texts have been decoded and the "voice" of Epicurean philosopher Philodemus is brought back again after 2000 years from rolled-up carbonized-Herculaneum-papyri. These findings provide new insights into Philodemus' work, going far significantly beyond a previous feasibility test-experiment which showed the benefit of XPCT applied to rolled-up papyri in imaging letters[5].

Thanks to the unprecedented contrast and quality of images provided by the new 'virtual-unrolling', Greek sequences of letters, words and, for the first time, full expressions and significant textual portions as well as a textual sign have been detected in unopened carbonized-Herculaneum-papyri.

This study paves the way to full interpretation of writings hidden in carbonized-Herculaneum-papyri, with the potential to expand the frontier of knowledge on ancient philosophy and literature. The experimental method applied is expected to impact across several disciplines, from materials and environmental science, to palaeontology and computerized archeological study techniques[6].

Since their discovery, numerous efforts have been made to open the carbonized-Herculaneum-papyri and read the precious works hidden inside and several destructive and non-destructive techniques were employed. Synchrotron XPCT, originally introduced mainly in bio-medical imaging, in the present study has been optimised and combined with versatile algorithms developed to deal with extracting distorted and corrugated papyrus.

This paper shows our success in "virtual-unrolling" two unopened papyrus rolls, i.e. *PHerc*. 375 and *PHerc*. 495, stored in Naples' National Library. A phantom sample, composed of high quality sheets of bare papyrus paper, was constructed and carbonized in order to determine the capacity and reliability of XPCT technique and the viability of the numerical algorithms. The phantom sample was earlier examined as an archaeological object in both geometry and material using XPCT. Subsequently, both papyrus rolls, *PHerc*. 375 and *PHerc*. 495, were measured through the same technique. In order to reveal the hidden text inside the original unopened papyri and the phantom sample, a new set of numerical algorithms for the 'virtual-unrolling' were developed 'in house'. These were tailored to adapt to the complex and diversified amorphous 3D deformed materials composing the rolls. The test performed on the phantom guided us, allowing to optimize and integrate the set of algorithms.

The result was a digital 'virtual-unrolling' of the unopened papyrus rolls with unprecedented resolution and contrast. Detailed tomographic images of the inner complex structure of the layers along the full longitudinal axis of the bookrolls were obtained. The optimized phase contrast 3D tomography distinguished areas of different densities: those composed of amorphous carbon-based ink were differentiated from that of the papyrus substrate, made of ordered carbon fibres. The 'virtual-unrolling' procedure, despite the complex organization in the inner portions of the scrolls, allowed to reveal Greek sequences of letters and words and, for the first time, read and decipher full expressions and significant textual portions as well as a *coronis*, a symbol used to mark the end of a textual section, a chapter or a book in classical papyri.

Different numerical procedures have recently been proposed to provide[7,8] a digital version of the text hidden inside rolled-up papyri starting from their non-invasive acquisition. Nevertheless the structure of carbonized-Herculaneum-papyri is quite different from that of a regular roll where a standard 'virtual-unrolling' algorithm is applicable. In recent years specific software has been developed[9] without achieving any reading of the text inside Herculaneum-papyri. Our success lies in combining XPCT with tailoring the basic principles of the 'virtual-unrolling' to the Herculaneum papyri and, consequently, virtually open, read and decode the largest portion of text, hidden inside unopened papyri, ever detected so far. This success is based on three main facts: i) the experience gained with the experiment on the phantom; ii) the optimal experimental conditions which provided, to the best of our knowledge, 3D images of the inner structure in carbonized-Herculaneum-papyri of the best quality ever obtained; iii) the very intricate internal arrangement of chaotic bundles of layers still presented sets of similarly bent, splayed and twisted layers. Once the XPCT provided the 3D volume of the bookrolls, 'virtual-unrolling' was performed. Upon partitioning the roll in subsets of similarly bent and twisted layers we were able to apply to each one 'flattening' procedures for each specific distortion of packed sheets.

Although the procedure is very elaborate and time-consuming, it is remarkable that the explorable partial area of the 'virtually unrolled' portion of the text was effectively unlimited. This adaptive procedure, allowing the complete unrolling of the rolls, is the propaedeutic tool for the first edition and interpretation of the works contained in unopened carbonized Herculaneum papyri.

For the first time the 'virtual-unrolling' allowed to unveil the largest portions of text ever read in unopened papyrus rolls. Figure 1a shows the sequence of the images obtained during the 'unrolling' procedure of *PHerc*. 375. From left to right of Figure 1a, an increasingly large portion of flat region appears, including an 11 mm large text of more than three lines. Its position inside the bookroll is shown in Figure 1b. Figure 2a shows the sequence of the images obtained from the 'unrolling' procedure of *PHerc*. 495. Figure 2b discloses the *coronis* with an exceptional image quality.

The revealed text was written on the papyrus side where the writing runs along the fibres (*recto*). Both rolls were ascribed on palaeographical grounds to Philodemus of Gadara, a remarkable philosopher and poet who studied in Athens before moving to Rome and Campania. In terms of textual extension, this assignment goes far significantly beyond that revealed in the previous XPCT study⁵. The handwriting was found to be a known typology in the Herculaneum library. As usual in classical antiquity, this was in capitals, with no word spacing, breathings, accents or lectional signs. The 'virtual-unrolling' allowed i) to reconstruct a complete Greek alphabet for each bookroll; ii) to detect several Greek sequences of letters and words, reported in Figure 3a-d and 3e-left; iii) to read for the first time, full expressions and significant portions of text in Figure 3c-e right-f-gleft and Figure 2; iv) to identify a textual sign (*coronis*), reported in Figure 3g right and Figure 2. These all are clearly distinguishable from the papyrus substrate. Using the 'virtual-unrolling' it was possible to perform the reconstruction of the largest portions of text ever detected in unopened papyrus rolls. Three distinct sequences of letters, i.e. ]ενερ, ]ψιc [ and ]κωc [ were detected in *PHerc*. 375. These are displayed from left to right in the top row of Figure 3 a; three words, i.e. τηρῆι (*tērēi*), περιε (*perie*) and π]ειcθεῖεν (*peistheien*), are shown in Figure 3b. In the case of *PHerc*. 495, the sequences ]επ[ (*ep*), ]cτ[ (*st*) and ]εχε[ (*eche*) are shown in Figure 3d. Furthermore, it was possible to isolate the word ]τείπομ[εν (*teipomen*) (left of Figure 3e) and the expressions ]εν γά[ρ (*en gar*) (right of Figure 3e) and ]τελη βημ[α (*telē bēma*) (Figure 3f). This last expression is formed by the second component stem of a compound adjective in -τελής, -ές or, less probably, the plural noun τέλη (from τέλοc, 'end' or 'purpose') and the noun βῆμα, 'tribune', in either the singular or the plural. The latter means the raised place from which orators spoke in a public assembly or in a lawcourt and is used in Philodemus' *On Rhetoric* (col. 56, 8 Sudhaus II; col. 264, 21 Sudhaus I). The *coronis* is clearly shown in Figure 3g right.

An example of the largest textual portions ever revealed inside unopened papyrus rolls is reported in Figure 3c, for *PHerc*. 375, and Figure 3g left, for *PHerc*. 495, together with a proposed textual reconstruction. In the former case, the verb ἐδέδ]οκτο might refer to a resolution taken by e.g. a political assembly. In the latter, the focus seems to be on either civil or political matters. It must be stressed that the different size and the deformation observed in some letters as well as some of the images is due to the distortion produced by the irregular surface of the papyrus substrate. Finally a *coronis*, namely a typical sign used by scribes in antiquity to signal specific articulations of the text, was revealed in *PHerc*. 495 (Figure 3g right).

Beyond the discovery of a large number of letters and texts, the 3D tomography revealed the first textual sign ever discovered in an unopened papyrus roll. In terms of spatial resolution and contrast, the present 'virtual-unrolling' permitted to fully exploit XPCT techniques.

The large number of letters and texts, revealed by the 'virtual-unrolling' made possible to analyze the handwriting within each unopened roll. Despite the distortions of the letters caused by the irregular surface of the layers, detailed paleographical description was achieved. From a comparative analysis with other Herculaneum papyri, both unopened rolls could be attributed to the Epicurean philosopher Philodemus of Gadara, bringing his 'voice' back after 2000 years. In particular, *PHerc. 495* is likely to represent a new unknown book of his twenty-books large treatise *On Rhetoric*[10]. Morover, our study revealed unexpected events historically experienced by the rolls.

These results demonstrate the potential of XPCT technique in reading inside rolled-up carbonized-Herculaneum-papyri. The investigation should enable the current knowledge of Hellenistic philosophy and classical literature to be expanded, with expected impact in fields such as papyrology, palaeography, classical philology and history of philosophy.

**Acknowledgements**

The authors wish to thank Luigi Nicolais (CNR), Glenn Most (SNS, Pisa), Daniel Koger (Lindsey Wilson College, Kentucky, US), G. Neville Greaves (University of Cambridge, UK) for valuable discussions and revisions of the present study, the ESRF Directorate for having granted the beamtime, the Biblioteca Nazionale 'Vittorio Emanuele III' of Naples, for lending us the samples (in particular, the officers Sofia Maresca and Vincenzo Boni), Eugenio Amendola (CNR-IPCB), for valuable assistance in the preparation of the containers for the samples, Carlo Ionta for technical assistance in data-analysis, Luigi Verolino (University of Naples Federico II) and Gaetano Campi (CNR-IC), for preliminary discussions.

**Author contributions**

A. C. proposed and coordinated the experiment, G. R. reconstructed the Greek texts and wrote the historical account, M. A. made the palaeographical analysis, R. B. coordinated the design and experimental measurements on the papyri puppet, R. G. A. and M. G. developed the concept of the papyri puppet and realized them. I. B. developed the algorithms for the 'virtual-unrolling' procedure and A. M. performed the tomographic reconstruction. A. B, P. C., F. C. and V. F contributed to the experiments and the data collection at the ESRF beamline, G. F., M. F., L. M. designed the figures. C. A., R. B., A. L., G. G. contributed to the theoretical discussion and the writing of the manuscript. All authors discussed the results and contributed to the redaction of the manuscript.

**FIGURE CAPTIONS**

**Figure 1:** *PHerc*. 375. **a,** virtual-unrolling; **b** textual portion.
**Figure 2:** *PHerc*. 495. **a,** virtual-unrolling; **b,** image of *coronis*.
**Figure 3: Sequences of letters, words, textual reconstructions and *coronis*, revealed in papyrus rolls through the 'virtual-unrolling'. a-c**, *PHerc*. 375 (top panel) and **d-g,** *PHerc*. 495 (bottom panel).

**Methods.**
*X-ray Phase Contrast Tomography imaging*

Standard X-ray tomography is based on absorption and it is a well known tool for imaging the internal structure of thick objects with hard X-rays. For low-absorption materials (like Carbon-based objects) small attenuation in the sample produces low contrast in the images. Thus standard absorption tomography is an unsuitable technique to discriminate details of similar densities in, for example, distinguishing carbon fibre-based papyrus foil from carbon-based ink used in writing on the papyrus. In the latter case, as shown in this study, a better contrast was achieved by imaging the phase modulation induced by the object in a coherent or partially coherent beam[3,4]. Several experimental approaches exist for detecting X-ray phase contrast. A simple yet effective phase-contrast method for hard X-rays is based on in-line imaging after free-space propagation.

*Experiment*

A series of experiments were carried out at ID17 of the European Synchrotron Radiation Facility (ESRF) in Grenoble (F), using a free-space propagation set-up. The imaging detector is a Fast Readout Low Noise (FReLoN) 2k charge-coupled device camera (CCD) connected with an X-ray optics determining an effective pixel size of 47 x 47 $\mu m^2$. The final spatial resolution in the images is about 100 x 100 $\mu m^2$. The field of view (FOV) was limited to about 1 mm (vertical) x 75 mm (horizontal). A double-silicon (111) crystal system was used to monochromatize the incident X-ray beam. Samples were located at about 10 meters from the CCD camera and they were rotated around the vertical axis parallel to the longitudinal axis of the rolls.

a) multi-energy scans were performed on the phantom sample in order to distinguish between absorption and phase contributions: incident X-ray beam was monochromatized at three different energies: 80, 52 and 30 keV. For each energy, 1400 radiograms were recorded for each scan, covering a total angle range of 180°, with acquisition time of 0.05 seconds per point. Results of these experiments allow to read and decipher letters and numbers previously written inside papyrus roll and to optimize the experimental conditions. Because of the flow of $CO_2$ during the combustion severe and diversified deformations developed within the bulk of the phantom papyrus roll. This complex inner structure was ideal to develop and test the new numerical algorithms for the 'virtual unrolling' and to tailor and optimize them to reveal the hidden writing accordingly. From the multi-energy experiment on the phantom sample it was also possible to record and to test the capability of distinguishing between the high signal due to the Pb in the ink from the low signal due to the other elements composing the ink.

b) The XPCT experiments on the PHerc.375 and PHerc. 495, were performed using a monochromatized incident X-ray beam with an energy of 73 keV. The choice of the energy is one of the essential conditions to achieve a suitable contrast in the final recorded images. The choice of the incident energy was selected following different guidelines: a) the experience gained with the phantom, which indicated the range of energies where contrast would have been higher; b) the measurement of the sample thickness which determined the minimum energy to be used; c) finally the optimizion was based on the maximization of the *contrast transfer function*[3] taking into account the fixed sample-detector distance and the pixel size of the CCD.

The two bookrolls, PHerc. 375 and PHerc. 495, were placed in cylindrical containers made of Plexiglas, specifically fabricated for the experiment, and safely mounted on a rotating plate at the sample position. About 2000 radiograms were recorded for each scan, covering a total angle range of 180°, with acquisition time of 0.05 seconds per point. Due to the limited FOV, in order to scan the whole scroll, several tomographic measurements were acquired at different vertical positions. The total data-acquisition time for each bookroll was about 5 hours.

Similar scans at low energy (52 keV) were performed on the *PHerc*. 495 and *PHerc*. 375 scrolls, in order to detect a possible fingerprint due to Pb present in the ink. The later image analysis did not reveal any trace of lead.

*Data processing*

In the first step of data analysis, the phase retrieval algorithm proposed by Paganin et al.[11] was applied to all projections of the tomographic scans using a modified version of the ANKAphase code[12]. Using a PyHST2 code[13], the recorded images were tomographically reconstructed. As a result a stack of 2D images representing the density distribution at different depths inside the sample was obtained. By means of a commercial 3D rendering software the 2D images of the stack were composed to build the whole 3D volume. The inner structure of the whole reconstructed rolls shows the various papyrus layers to be diversely bent and twisted at the different depths inside the volume.

*Image treatment*

3D rendering and segmentation of the impurities were carried out using the Visage Amira 6 software (Visualization Sciences Group Inc.)[14] and VolView[15] (see SI). The 3D structure of the papyri was obtained through a sort of segmentation procedure. The images were generated by associating colors to the different gray-scale ranges in the 3D model.

*Virtual Unrolling*

In the last years virtual unrolling problem using volumetric scanning in virtual restoration and preservation of ancient artefacts like parchments or papyri has been an extremely active area of research. The use of X-ray Computed Tomography (CT) and X-ray Micro Computed Tomography for the data digitalization has encouraged the development of restoration algorithms. Different numerical procedures have been proposed to provide a digital version of the text hidden inside rolled-up scrolls by starting from non-invasive acquisition.

The procedure of virtual unrolling can be divided into three main steps: volumetric scanning, segmentation, layered texture generation and restoration.

**Volumetric scanning:** In this stage the experimental data are acquired and their digitalization is performed.

**Segmentation:** The process of segmentation is used in order to segment a voxel set into two categories: material voxels versus empty space.

**Layered texture generation:** This is the stage where the surface modelling is performed.

**Digital restoration:** The surface flattening and unrolling can be interpreted as an isometric mapping (i.e. it preserve distances, which minimises text distortion in the parchment) from 3D to 2D images.

To solve these problems different approaches were proposed. One of the most promising is described by Seales and colleagues[7, 9]. The authors have developed software that integrates functions of flattening and unrolling based on mass-spring surface simulation. The algorithms proposed by O. Samko et al.[8] allows to solve the problem of touching points between adjacent sheet layers.

The internal structure of the Herculaneum papyrus is quite different from the regular roll where a simple 'virtual unrolling' was applicable. However the very intricate internal arrangement of chaotic bundles of layers still presented sets of similarly bent, splayed and twisted layers. The central part of the scroll is better preserved and the layers of papyrus are separated and loosely rolled up. The outer part of papyrus is tightly scrolled and stacked. It is evident that the virtual revealing of the text requires different approaches and has its own challenges. From the experiment performed on the prototype we learned that the letters can be deformed due to thermal and pressure impact and due to irregular surface of the fibres. *A priori* it is not possible to know the letters aspect and even whether the text is present in the analysed portion of rolls. In principle the text could be destroyed or could not appear due to the wrong position in the papyrus foil.

We performed semi-manually procedure to reveal the hidden text based on computed algorithms exploiting Matlab® codes ImageJ macros and different available commercial computer software. Once the phase contrast tomography provided the 3D volume of the bookrolls, to reveal the Greek text hidden inside, a multistep procedure has been settled:

1. Accurate optimization of the range of grey-levels has been done to individuate the inner structure of papyri with preserving all information about text.

2. Selection of areas with packaging of sheets with similar orientation.

3. Flattening of packed sheets, using different strategies including transformation of coordinates.

4. Filtering and segmentation of individual sheets in 2D slices.

5. Surface modelling.

6. Segmentation based on threshold to reveal the text

**FIGURE 1**

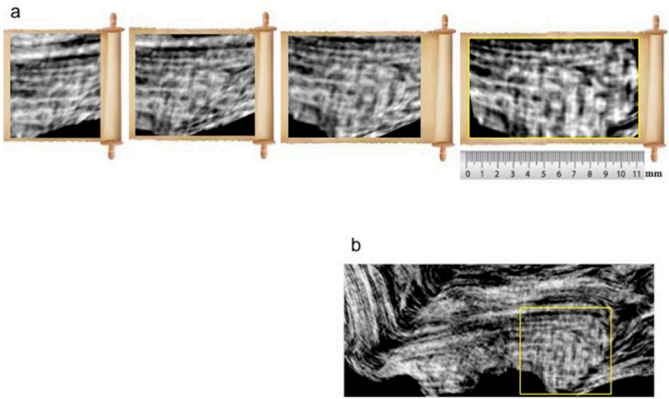

**Figure 1:** *PHerc*. 375. **a,** virtual-unrolling; **b** textual portion.

**FIGURE 2**

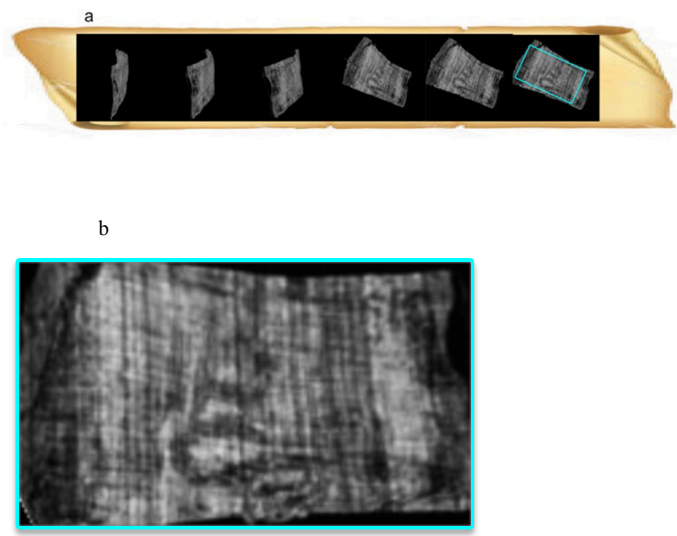

**Figure 2:** *PHerc*. 495. **a,** virtual-unrolling; **b,** image of *coronis*.

**FIGURE 3**

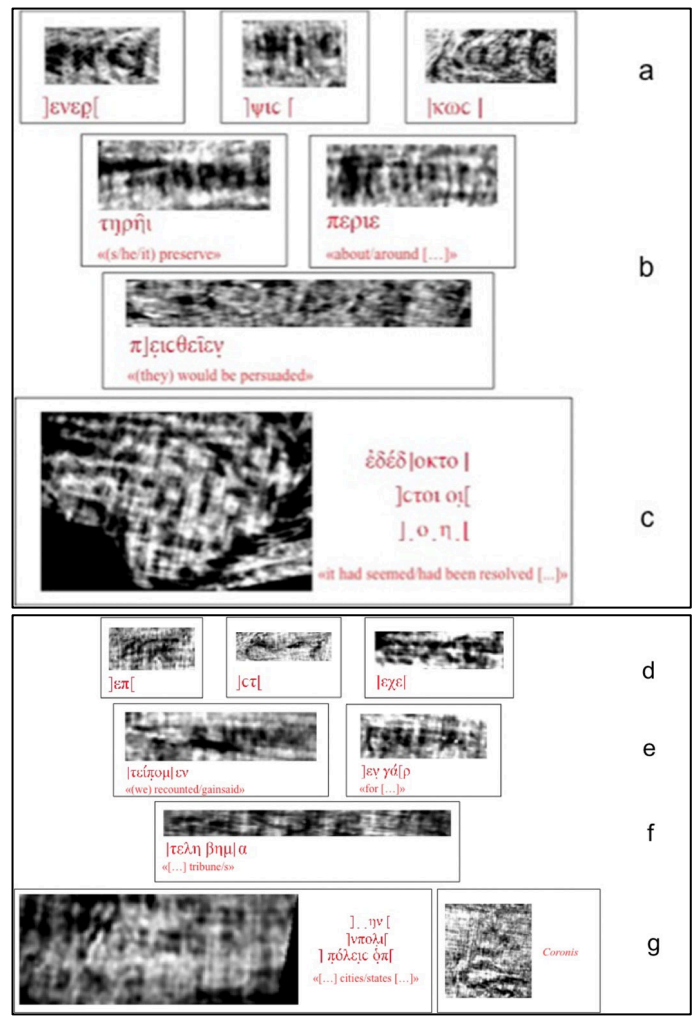

**Figure 3: Sequences of letters, words, textual reconstructions and *coronis*, revealed in papyrus rolls through the 'virtual-unrolling'. a-c**, *PHerc*. 375 (top panel) and **d-g,** *PHerc*. 495 (bottom panel).